# MAGNETIC FIELD STRUCTURE NEAR THE GALACTIC PLANE

**Andreasyan R.[1], Balayan S.[2], Movsisyan V.[3]**
1; 2; 3 Byurakan Astrophysical Observatory, 0213, Armenia

**Abstract**

We bring the two color maps for the plane component magnetic field of our Galaxy in coordinates of (R;l) and (DM;l). It was shown that magnetic field has reversals of the direction in neighbor spiral arms, in agreement with known models for the Galactic magnetic field. For the Sagittarius spiral arm region there is, however, some disagreement with standard magnetic field models. The major discrepancy is the fact that in the Sagittarius arm region the magnetic field in Southern hemisphere of the Galaxy, have opposite direction to the field of the Northern hemisphere. We think that the Sagittarius spiral arm, at last the magnetic spiral arm in this region is not symmetric to the Galactic plane, and is located mainly in Northern hemisphere.

## 1. Introduction

The activity of cosmic objects is associated in most cases with the presence of magnetic fields of different scales, natures and strength. It seems to be very important to take into account of magnetic field distribution in galaxies, and partially in our Galaxy, when we study the formation and evolution of galactic structural features as well as a whole morphology of optical and radio galaxies and quasars. The large-scale magnetic field in our Galaxy was found in 1950-th, and till now is studied hardly using all available methods based on the analyses of optical and radio polarization data and measurements of Rotation Measure (RM) of extragalactic radio sources and pulsars etc. It was shown that three classes of model are viable for the large-scale structure of magnetic field in the disk of Galaxy: 1) A bisymmetric spiral (BSS), in which the field direction reverses from arm to arm. 2) An axisiymmetric spiral (ASS), with two field reversals inside the Solar Circle. 3) A concentric ring model.

Men et all (2008) analyzing the pulsar data show that none of these models appears to perform significantly better than the others. From this they conclude that the large-scale interstellar magnetic field in the Galactic disk has a more complex pattern than just circular, axisymmetric, or bisymmetric.

In the recent study by Indrani & Despande (1998), Han et all. (2006) a model was suggested involving a magnetic field with the spiral structure lying in the inter arm regions. Observations by Beck & Hoernes (1996) show that in other galaxy (NGC 6946) also the magnetic spiral structure lays in the inter-arm region.

Observations of radio polarization of many edge-on galaxies can be explained including the Halo component of magnetic fields (For example: NGC 4631 (Hummel, Beck & Dahlem 1991); NGC 253 (Beck et. al.1994; Heesen et all. 2009); NGC 4666 (Dahlem et al.1997); M33 (Tabatabaei, et. all 2007); M51 (Fletcher et all. 2010)). Heesen et all. 2009 show that the large-scale magnetic field NGC253 was decomposed into a toroidal axisymmetric component in the disk and a poloidal component in the halo. The poloidal component shows a prominent X-shaped magnetic field structure centered on the nucleus, similar to the magnetic field observed in other edge-on galaxies. For the galaxy M51 (Fletcher et all. 2010) was derived two components for the regular magnetic field, the disc is dominated by a combination of azimuthal modes, m=0+2, but in the halo only an m=1 mode is required to fit the observations.

The two component magnetic field of our Galaxy was studied by Andreasyan & Makarov (1988, 1989). It was suggested that the distribution of RMs of pulsars and extragalactic radio sources are consistent with the model, when the plane magnetic field component of spiral arms is imbedded in the Halo dipolar magnetic field of Galaxy, which has opposite directions above and below the Galactic plane. Han et al. (1997) (1999) obtained a similar result for the Halo magnetic field of the Galaxy. It was also estimated the vertical component of magnetic field to be B=0.37 μG, directed toward the North Galactic pole.

In the recent paper Moss et all. (2010) show that by including a galactic wind, in standard galactic dynamo models can by obtained two component magnetic field configurations, approximately even in the disc and odd in the halo.

So, in spite of a large number of papers studying the structure of Galactic magnetic field, there is no generally accepted model. We think it is very important to study the z (distance from the Galactic plane) dependence of Galactic magnetic field.

### 2.1 The mapping of Galactic magnetic field (the method)

In the present paper we construct the two color maps for different layers (divided by the z coordinate) of the Galactic magnetic field using all available rotation measure data of pulsars, which gives a possibility to study the 3-dimensional distribution of this magnetic field. Since the time of our earlier studies, much more data has become available for this investigation, particularly for more distant pulsars (Weisberg et al. 2004; Han et al. 2006; Noutsos et al. 2008). The total number of pulsars with known values of the rotation measure -RM, now is ~600. We use this improved database for our study.

It is well known that pulsars are strongly concentrated to the galactic plane, and there are not so many pulsars in the Halo region. Therefore here we study the magnetic field of plane component (spiral arm region of the Galaxy) and receive the maps for distribution of magnetic field in the galactic plane. It is known that Faraday rotation measure (RM) and dispersion measure (DM) for pulsars are functions of their distances R from the Sun, and are given by the formulas

$$RM = \alpha \int^R n_e B_L dL, \quad (\alpha = 8.1 \cdot 10^5) \quad (1)$$
$$DM = \int^R n_e dL, \quad (2)$$

where $B_L$ is the component of the magnetic field along the line of sight (in Gs), R- the distance of pulsar from the Sun in pc, $n_e$ is the electron density ($cm^{-3}$), and the integral is taken over a distance L (pc). Dispersion measures and rotation measures of pulsar are known from the observations (DM - practically for all pulsars and RM – for 595 pulsars). Equation (1) and (2) yield:

$$\langle B_l \rangle = (1/\alpha)(RM)/(DM), \quad (3)$$

where $\langle B_l \rangle$ is the magnetic field strength averaged along the line of sight

Using the equations (1) and (2) for the given direction l (l – is the galactic longitude) can be received:

$$B_l(R)n_e(R,l) = (1/\alpha)d(RM)/d(R) \quad (4)$$

and

$$B_l(DM) = (1/\alpha)d(RM)/d(DM) \qquad (5)$$

where $B_l(R)$ is the line of sight component of magnetic field strength at the point with a distance R from the Sun (unlike to averaged value of $\langle B_l \rangle$). and $B_l(DM)$ is the line of sight component of magnetic field strength at the point with a given value of DM. It means, that using the RM(R) and RM(DM) dependences for a given direction, it is possible to find (using the formulas (4) and (5)) $B_l(DM)$ for each value of DM, and $B_l(R)n_e(R,l)$ for each value of R. We can find the RM(DM) or/and RM(R) dependence for all directions in the plan of Galaxy using averaging procedure similar to one presented in Andreasyan et.al (2006). That is, we use the method, when the coordinates (l;DM) of the center of averaging region (where - l is the galactic longitude, DM – the dispersion measure), is changing smoothly in the plane of (l;DM). So we can find the dependence of average values of RM from the average value of DM in every direction, and from the formula (5) find the $B_l(DM)$. This is true also for the RM-R dependence, and from the formula (4) we find the $B_l(R)n_e(R,l)$. In fact we are solving the inverse problem finding the internal distribution of physical parameters using the line of sight integrals of these parameters. In the result of these calculations we can construct the 2-dimensional maps for plane component of Galactic magnetic field in coordinates (l;DM), or (l;R).

## 2.2 The mapping of Galactic magnetic field (the results)

The results of calculations are given with two color maps of $B_l(R)n_e(R,l)$ and $B_l(DM)$ (fig.1 and fig.2). The program gives a possibility to change many parameters during the calculations. For example: the minimal number of pulsars in the averaging regions, to change the studying layers with z coordinate for the study of magnetic field changes in third direction (z=R·Sin(b), where b- is the galactic latitude), to show or not the pulsars on the maps, et cetera. In the fig.1 we have the distribution of $B_l(R)n_e(R,l)$ in the galactic plane (l;R). The Sun is located in the center of the distribution. The center of the galaxy is directed to the right (green point), and the galactic longitude l increases opposite to the clockwise. The distance of the sun from the center of Galaxy is accepted 8.5 kpc. The blue color indicates that in this region magnetic field component is directed to the observer (RM>0), and the red color we use for the magnetic field component directed from the observer (RM<0). As dense is the color, as large is the value of $B_l(R)n_e(R,l)$ (the value of this is indicated on the bar). On the picture we have the distribution of $B_l(R)n_e(R,l)$ for all pulsars with known RM (-1800<z<1800pc, fig.1a) and for the plane component of magnetic field (-400 <z<400pc (fig.1b), -400<z<-50pc (fig.1c) and -50<z<400pc (fig.1d)). The restriction |z|<1800pc comes from the catalogues of pulsars (no pulsars with z>1800pc), and z=-50 we choose as a best border for the changes in z direction (See also the section 3). On the map with -400 <z<400pc and -1800<z<1800pc we see the reversals of magnetic field directions from one spiral arm to another, what is consistent with the results of previous studies (for example, see Han et al. 2002, 2006). In the maps with -400<z<-50pc and -50<z<400pc (fig.1c and 1d) we bring the distribution of $B_l(R)n_e(R,l)$ for Southern and Northern hemispheres of the Galaxy. We see significant differences in these two distributions. The main difference in the pictures for Southern and Northern hemispheres is the magnetic field distribution in the direction of Sagittarius spiral arm (l ~ 55°). The very strong and homogeneous magnetic field of Sagittarius spiral arm, directed to the observer (blue color), appears only in the Northern hemisphere. This result is consistent with the results of Andreasian et al (2003). On the fig.1 there can be found other large scale features also, that will be discussed in other papers, but here we want to focus on one methodical point.

We must note that the results, obtained from the fig.1, for Plane component of Galactic magnetic field depends strongly from the method of estimation of pulsars distances. It is obvious, that these results reflect the spiral arm model of electron density distribution (we use the results of Taylor & Cordes, 1993), used for the estimation of pulsar distances. It is the reason, that for the detail investigation we use also the distribution of $B_l(DM)$, where we use only the observational data, and don't use any model of galactic electron density. We bring here, for example, one of these distributions (fig.2). In picture 2, as one of coordinates, we use the dispersion measure DM instead of distance R from the Sun. The galactic longitude increases opposite to clockwise (the center of the galaxy is directed to the right). From the picture we see, that the maps for pulsars with $-50<K<50$ ($K=DM \cdot Sinb$ pc·cm$^{-3}$) or/and $-20<K<20$ (fig.2a and 2b) are similar to fig.1a and fig.1b with reversals of the direction of magnetic fields in neighbor spiral arms. For the pulsars with $-20<K<-1$ and $-1<K<20$ (fig.2c and 2d) also we see the same difference as in fig.1c and 1d. The main difference in the pictures for Southern and Northern hemispheres, is the magnetic field distribution in the direction of Sagittarius spiral arm $l\sim55^o$. The strong magnetic field of Sagittarius spiral arm appears only in the Northern hemisphere. It must be noted that the fig.2 gives a new method to find the magnetic spiral arms in terms of (l, DM). These spiral arms can coincide with known spiral arms (see, for example, Georgelin & Georgelin 1976, or Taylor & Cordes 1993), or not. From the comparison of our magnetic field maps (fig.1 and fig.2) with the spiral arm structure of Galaxy we see that the magnetic field in the direction of Sagittarius has positive sign (blue color) and in the direction of Carina, which is the continuation of the Sagittarius spiral arm to the another side of the Sun, the magnetic field has negative sign (red color). It means that in all Sagittarius-Carina spiral arm we have one large-scale magnetic field with a little pitch angle. The same is true for the Local Orion arm, but with the magnetic field of opposite direction. The magnetic field of the Perseus spiral arm is directed in the same way as the magnetic field of the Local Orion arm. It is in consent with the model, where the Local Orion arm is proposed to be as the branch of the Perseus spiral arm. These results for the plane component of Galactic magnetic field were mainly known also from the previous studies of many authors, but without of the dependences of magnetic field from the z coordinate. In section 3 we bring one of these dependences for the Sagittarius spiral arm.

### 3. Magnetic field in the Sagittarius spiral arm region

We reexamine here some results for the Sagittarius Spiral arm from our early paper (Andreasian et al. 2003) using new and more RM data of pulsars. The data of pulsars from Sagittarius arm region ($l\sim55^o$) were used to construct a two dimensional plot of the distribution of the signs of RM for the pulsars (Fig.3). In the direction $40^o<l<80^o$ there are 60 pulsars and for the construction of fig.3 were used the data of these objects. This region can be chosen also from the study of pictures 1 and 2. The abscissa of Fig.3 shows the distances of the pulsars from the Sun in the galactic plane (R). The ordinate shows the distances of the pulsars from the galactic plane (z). The pulsars are indicated by circles: solid if the sign of RM is positive, and open if RM is negative. The distribution of the signs of RM for pulsars manifests the following main behavior:
a) The pulsars with negative RM are mainly located in Southern hemisphere: b) The pulsars with positive RM are located mainly in Northern hemisphere, except for a few pulsars with positive

RM in the Southern hemisphere but very close to Galactic plane. The distance from the Galactic plane to some of these pulsars is less than 50 pc.

Fig.4 is a plot of the distribution of rotation measures of the pulsars with respect to their dispersion measures. This figure shows that: a) Up to DM=30-40 pc. cm$^{-3}$, negative RM values of Southern pulsars are almost linearly decreasing, while beginning with DM = 40 pc. cm$^{-3}$ there is not a linear RM-DM dependence. The absolute values of RM for these pulsars are extremely small and do not correspond to their DM or to their distances. b) For the Northern hemisphere pulsars with positive RM there is a significant linear dependence up to DM=250 pc. cm$^{-3}$. It is interesting to note that 4 pulsars (B1929+20, B1930+22, B2002+31, B2011+38) on the right lower corner of fig.4 with DM>200 pc. cm$^{-3}$, but exceptionally small value of RM, are relatively younger, with ages of less than a million years. As it was shown by Pinzar & Shishov (2001), large values of DM (100-150 pc. cm$^{-3}$) for young pulsars may be caused by HII regions (supernova remnants) coupled to them. Then these pulsars may seem to be nearby objects, which may explain their relatively low RM. Thus, if we shift them to the left on fig.4 by roughly 100-150 pc. cm$^{-3}$, then we can obtain better linear relationship between RM and DM for the Northern pulsars.

**4 Summary and conclusion**

The main results of this study are the maps of the plane component of Galactic magnetic field with reversals of the direction in neighbor spiral arms, basically in agreement with standard models for the Galactic magnetic field. There is, however, some disagreement between our results and the standard magnetic field models. The major discrepancy is the fact that in the direction of the Sagittarius arm the magnetic field in the southern hemisphere of the Galaxy, while very weak, is in the opposite direction from the field of the northern hemisphere. The borderline where the magnetic field changes direction is parallel to the Galactic plane and lies in the southern hemisphere roughly at a distance z~50 pc from the Galactic plane (see also the Fig1 and Fig2.). Since it is difficult to conceive that the magnetic field could change direction in a thin layer of the spiral arm, we assume that the arm of Sagittarius in the region $40°< l <80°$ lies entirely to the north of this borderline, i.e., the borderline at which the direction of the large scale magnetic field changes, lies outside the Sagittarius spiral arm. We think that the Sagittarius spiral arm, at last the magnetic spiral arm is not symmetric to the Galactic plane. The negative RM of pulsars in the southern hemisphere within $40°< l <80°$ may be attributed to a contribution from the magnetic field of the Galactic halo.

The study of obtained magnetic field maps, their dependence from the z coordinate, shows that there are other details that also are distributed not symmetrically to the Galactic plane, which will be discussed in other paper.

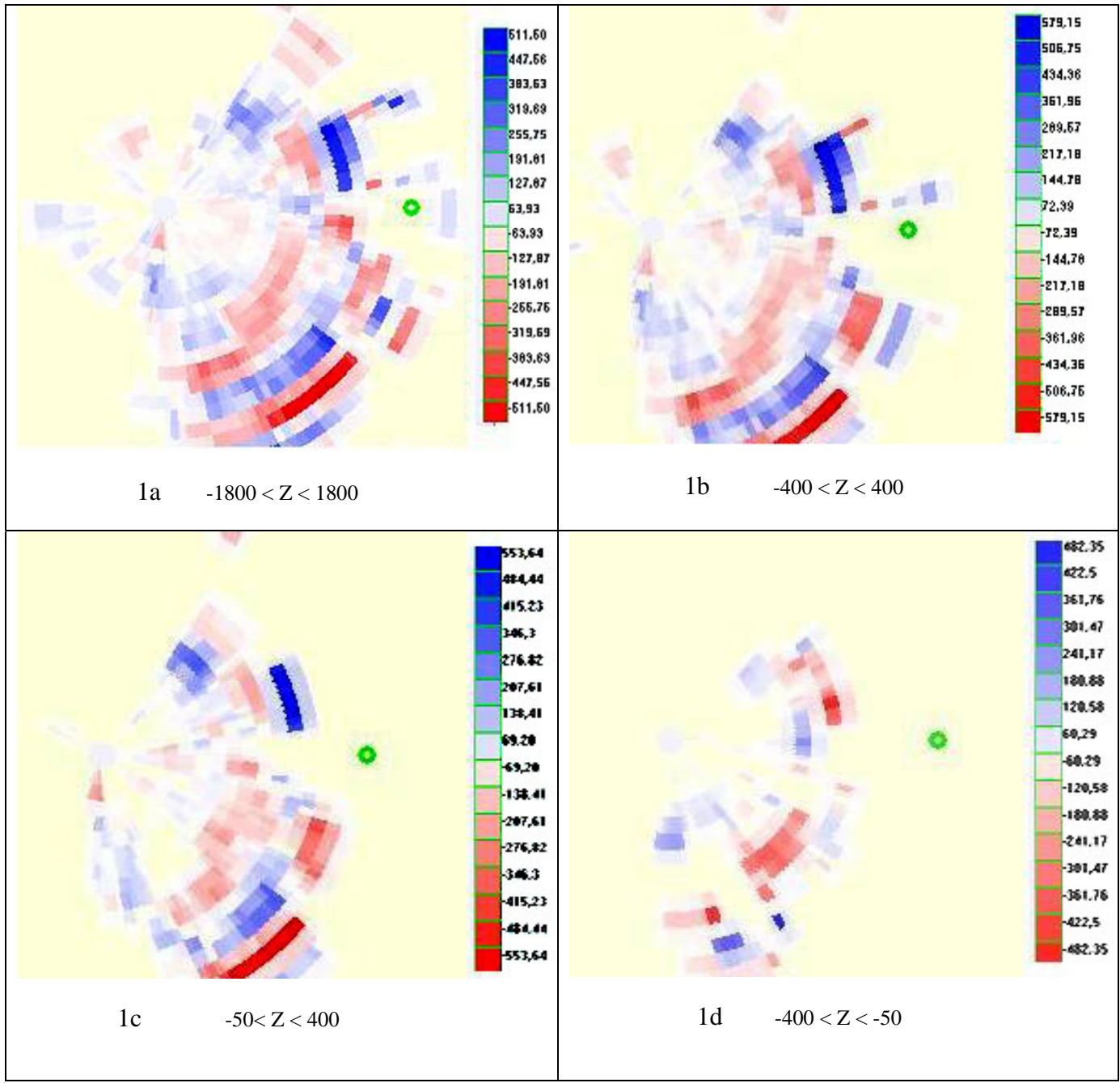

Fig.1 The distribution of $B_L(R)n_e(R)$. in the galactic plane $(R;l)$.

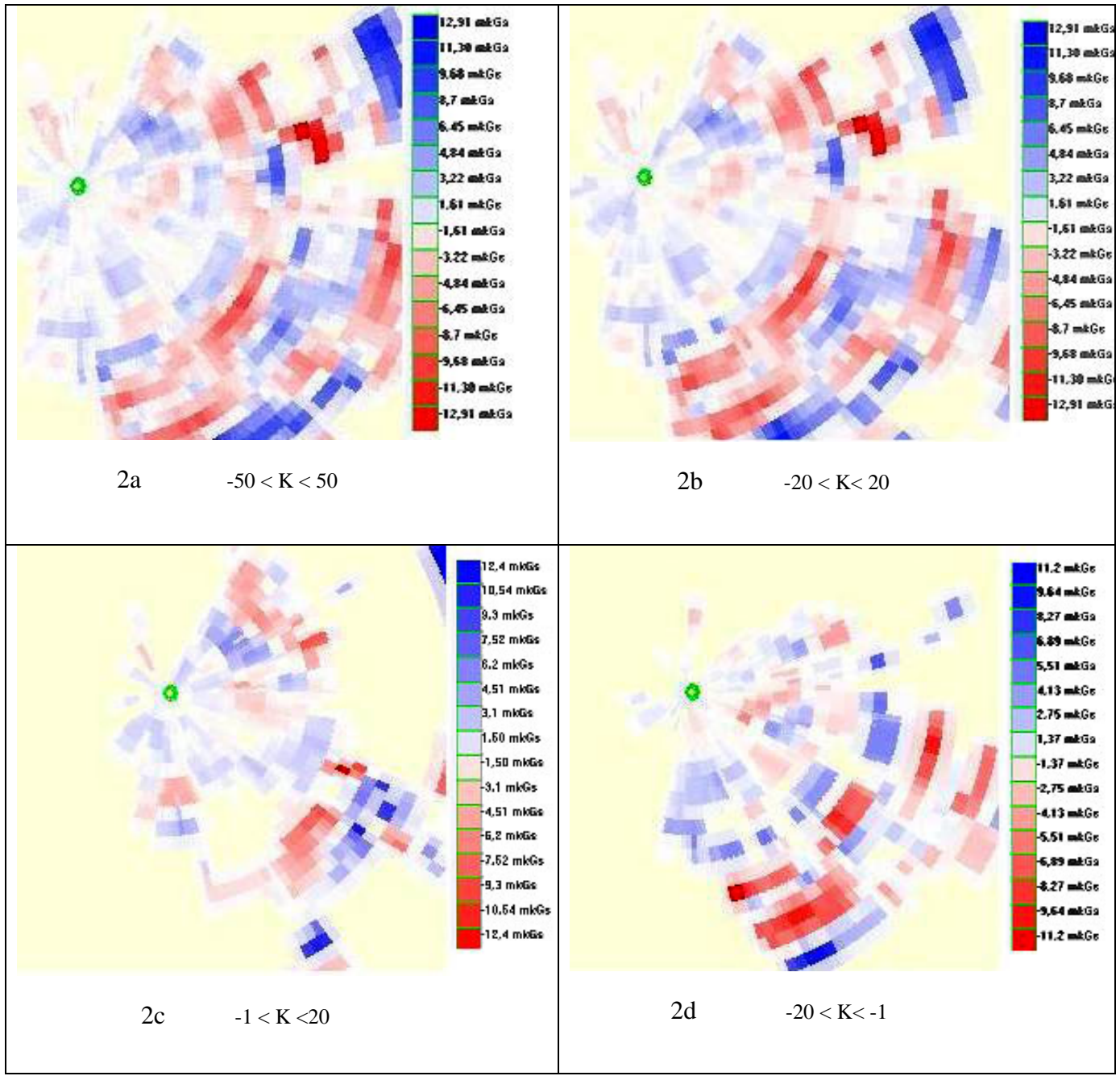

Fig.2 The distribution of $B_L(DM)$ in the galactic plane (DM;l).

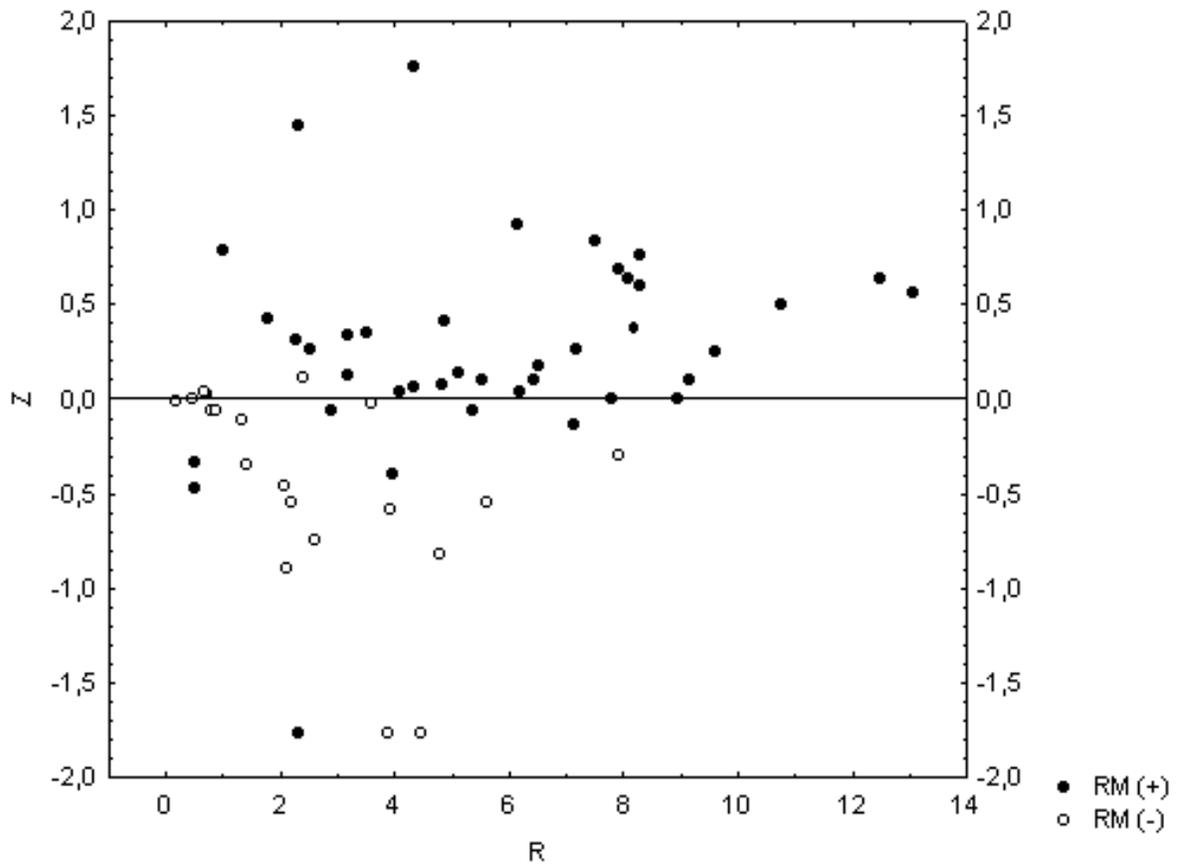

Fig.3 The distribution of the signs of RM for the pulsars in the direction $40°<l<80°$, The horizontal and vertical axes are accordingly the distance from the Sun and from Galactic plane.

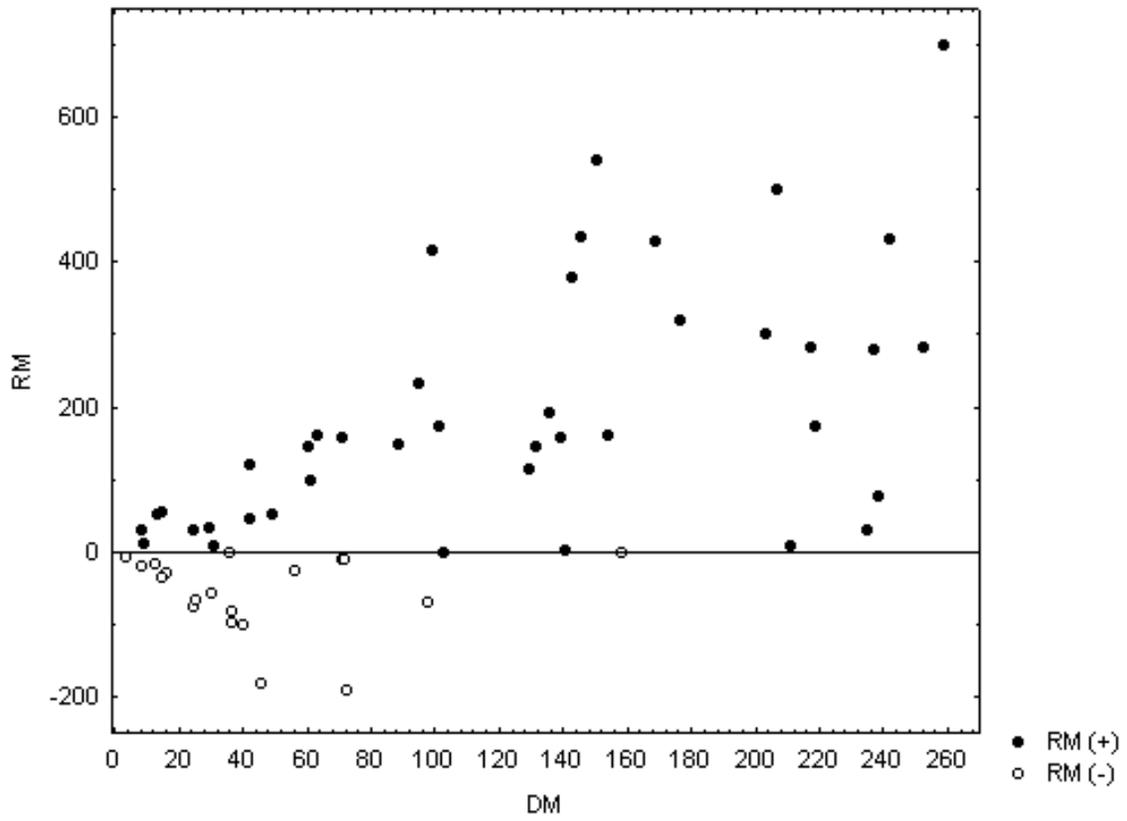

Fig.4 The distribution of RM of the pulsars with respect to their DM.